\documentclass{webofc}
\usepackage[varg]{txfonts}   
\usepackage{lineno}
\usepackage{amsfonts}
\usepackage{hyperref}

\begin{document}

\title{Quarkonia as probes of initial and final states in small systems with ALICE}
\vspace*{-0.1 cm}
%
%

\author{\firstname{Subikash} \lastname{Choudhury} for the ALICE Collaboration \inst{1}\fnsep\thanks{\email{subikash.choudhury@cern.ch}} 
}

\institute{ Saha Institute of Nuclear Physics, 1/AF, Bidhan nagar, Kolkata 700064, India}

\abstract{%
The multiple parton$-$parton interactions (MPIs) are an important element to describe the observed collective flow and strangeness enhancement in high multiplicity pp and  p$-$Pb collisions, the so-called small systems. 
At LHC energy, MPIs affect both soft and hard scales of the event. Since quarkonium production involves both, it can be an excellent tool to understand the role of MPIs in small systems. Study of multiplicity dependent quarkonium production provides an indirect probe of MPIs in hadronic collisions. Also, relative production of excited-to-ground quarkonium states as a function of multiplicity are sensitive to final state effects. Cross section measurements of different quarkonium states are also important to understand their production mechanisms.
In these proceedings, first preliminary results of $\psi$(2S)-over-J/$\psi$ cross section measurements at mid and forward rapidity (\textit{y}) from Run 2 and Run 3 pp collisions at $\sqrt{s}$ = 13 and 13.6 TeV, respectively will be presented. Preliminary results on $\Upsilon$(nS) states (n = 1,2) cross sections and final multiplicity dependent measurements of excited-to-ground states quarkonium yields, at forward \textit{y}, in pp collisions at $\sqrt{s} =$ 13 TeV will be shown. In addition, few performance plots from Run 3 in the quarkonium sector will be presented and discussed.
}

\maketitle

\section{Introduction}
\label{intro}
Quarkonia are bound states of heavy quark$-$antiquark pairs, like charmonia (c$\mathrm{\bar{c}}$) and bottomonia (b$\mathrm{\bar{b}}$), which are produced in the early stages of hadrnoic collisions.  Because of their large masses, charm and bottom quark$-$antiquark pairs are produced via high momentum transfer processes (hard scatterings),  which are described well by perturbative Quantum Chromodynamic (QCD) calculations. The formation of bound states however, involves soft-QCD processes. Thus, quarkonia production is sensitive to both perturbative and non-perturbative aspects of QCD. 
Measurements related to quarkonium production in small systems, where quark gluon plasma (QGP) formation was historically not expected, can also serve as a suitable reference for heavy-ion collisions. However, several intriguing QGP-like features such as collective flow~\cite{ALICEBA_PLB726_2013} and strangeness enhancement~\cite{ALICEJA_NP13_2017}, have been observed in small collision systems in high multiplicity events. A scenario based on MPIs in a single hadron$-$hadron collision describe some of these findings reasonably well. An indirect approach to probe MPIs in hadronic collisions is to study quarkonium production as a function of multiplicity. Besides that, excited-to-ground state quarkonium yields as a function of multiplicity can reveal effects from other mechanisms at play in the final state, such as the dissociation by the surrounding comoving particles of loosely-bound excited quarkonium states~\cite{EF_PLB749_2015}.

In these proceedings, preliminary results of $\psi$(2S)-over-J/$\psi$ cross section ratios are presented for the first time at midrapidity (|\textit{y}| < 0.9) in pp collisions at $\sqrt{s} =$ 13 TeV as well as at $\sqrt{s} =$ 13.6 TeV for both mid and forward rapidity (2.5 < |\textit{y}| < 4.0) using for the first time Run 3 data. The transverse momentum (\textit{p}$_{\mathrm{T}}$)- and \textit{y}-differential cross sections of $\Upsilon$(nS) states (n = 1,2) at forward \textit{y}, and self normalized excited-to-ground $\Upsilon$(nS) states yield ratios as a function of the charged-particle multiplicity at midrapidity are also shown in pp collisions at $\sqrt{s} =$ 13 TeV. Finally, few performance plots are presented that portray future prospects for quarkonium measurements from Run 3 data in ALICE.

\section{Experimental setup}
\label{ALICE}
With the ALICE detector setup~\cite{ALICEBA_IJM29_2014}, quarkonium measurements can be done at mid- and forward- \textit{y} in dielectron and dimuon decay channels, respectively. At mid-\textit{y}, quarkonium measurements are done in the dielectron decay channel using the Inner Tracking System (ITS), the Time Projection Chamber (TPC), and the Transition Radiation Detectors (TRD). 
At forward \textit{y}, the muon spectrometer is the main detector used for quarkonium measurements in the dimuon decay channel. To measure charged-particle multiplicity at midrapidity, the innermost layers of the ITS, which are Silicon Pixel Detectors (SPD), are used. For Run 3, ALICE has upgraded its ITS and TPC, and has installed a new Muon Forward Tracker (MFT)~\cite{ALICEMFT_TDR_2015} to reconstruct charged particles in the muon spectrometer acceptance upstream of the absorber. With the MFT, secondary vertexing can be done, which will allow separation of prompt and non-prompt quarkonia at forward \textit{y} for the first time in ALICE. Also, a continuous read-out mode data acquisition was introduced to collect larger integrated luminosities.

\section{Results}
\vspace*{-0.5cm}
\label{sec-2}
\begin{figure*}[hb]
\centering
\includegraphics[scale=0.27]{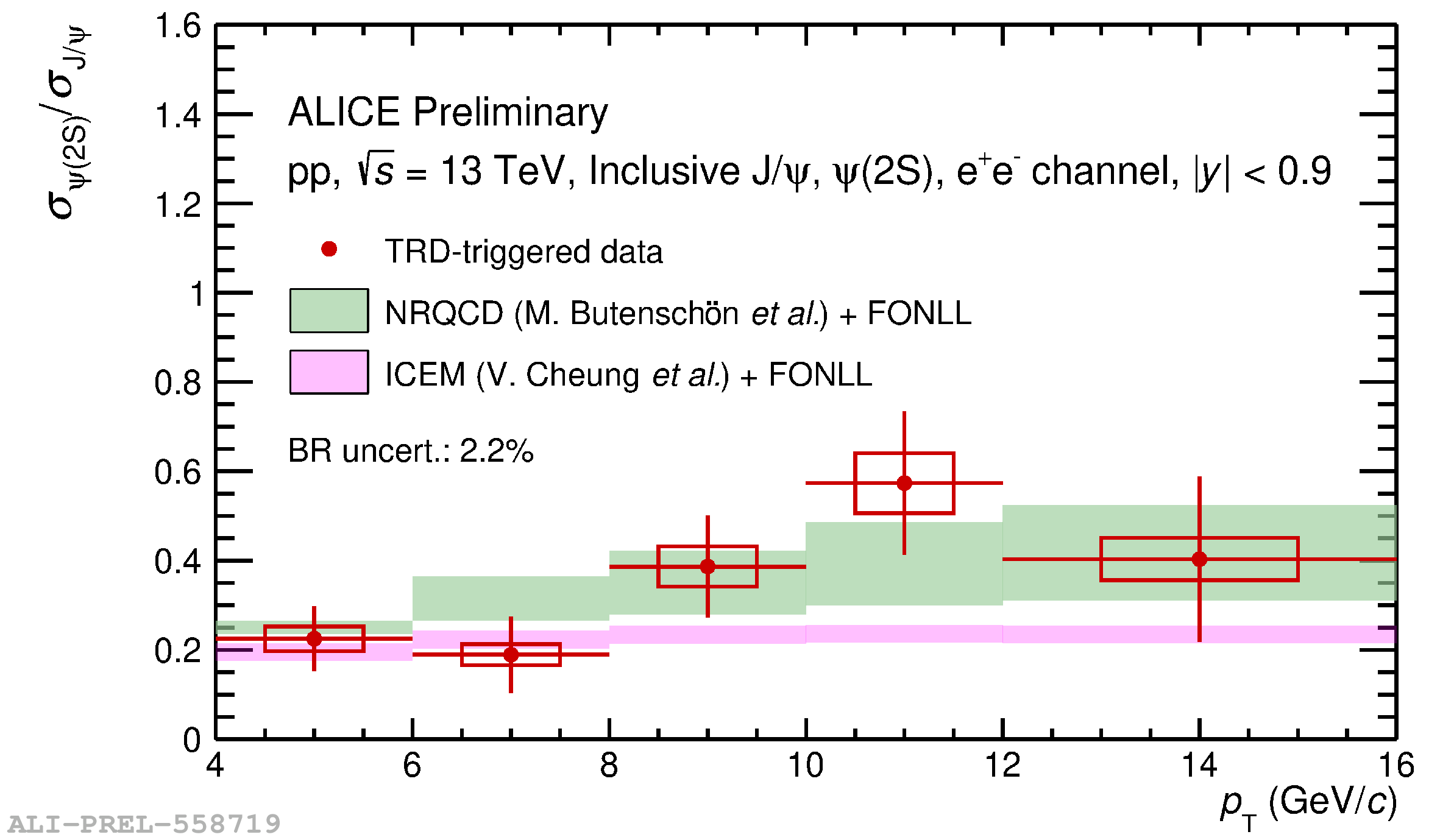}
\includegraphics[scale=0.25]{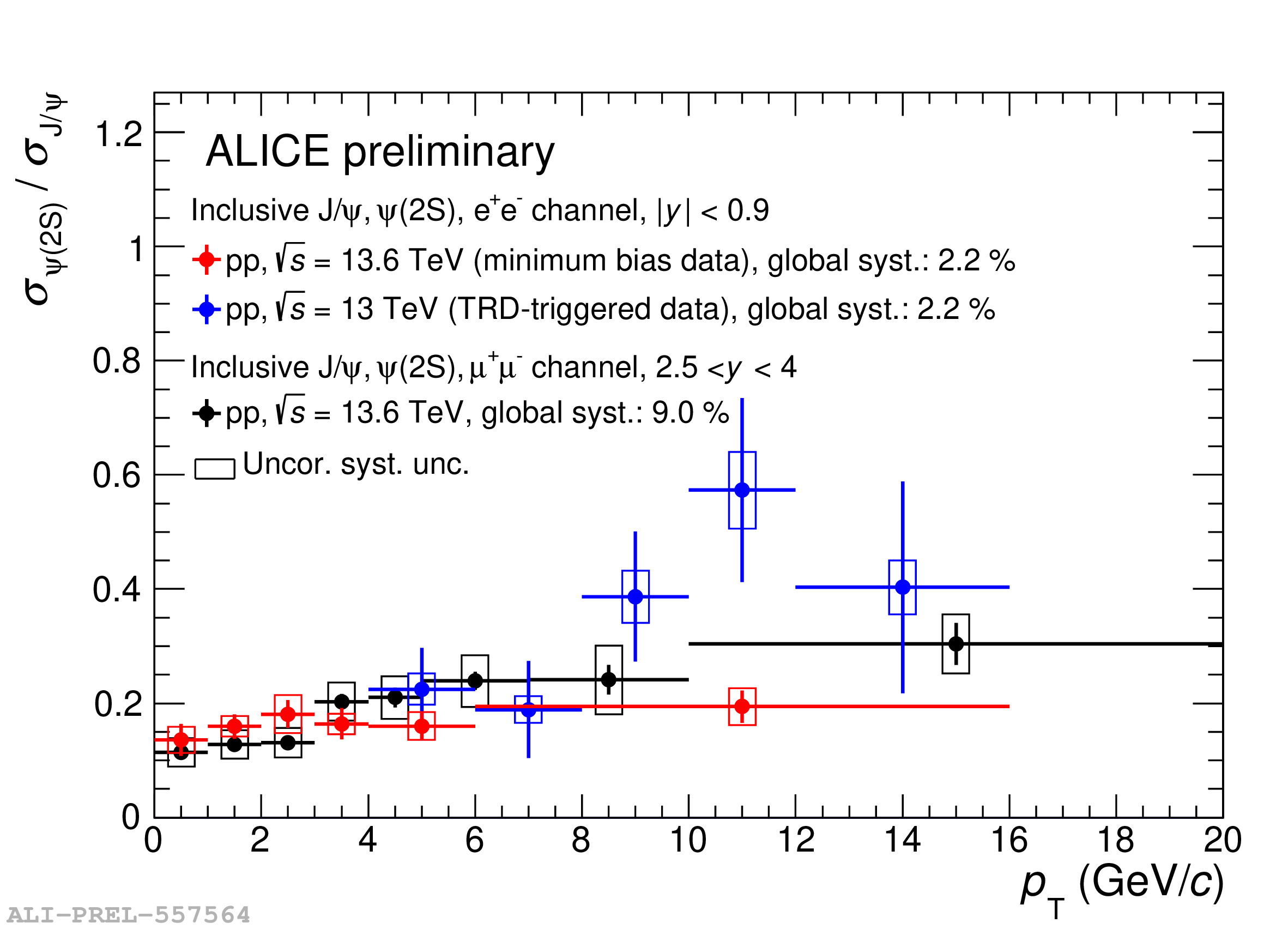}
\caption{(Left) Inclusive $\psi$(2S) / J/$\psi$ cross section ratios as a function of \textit{p}$_{\mathrm{T}}$ in TRD-triggered pp collisions at $\sqrt{s}$ = 13 TeV. Results are compared with NRQCD~\cite{MB_PRL106_2011} and ICEM~\cite{VC_PRD98_2018} model calculations. (Right) First results of the \textit{p}$_{\mathrm{T}}$-dependent $\sigma_{\psi(\mathrm{2S})}$ / $\sigma_{\textit{J}/\psi}$ ratios at mid and forward \textit{y} from minimum-bias Run 3 data in pp collisions at $\sqrt{s}$ = 13.6 TeV. The Run 3 data are compared with the Run 2 midrapidity results from TRD-triggered pp collisions at $\sqrt{s}$ = 13 TeV.  }
\label{fig-1}       
\end{figure*}
For the first time, ALICE has measured inclusive $\psi$(2S) yield at midrapidity using TRD-triggered Run 2 pp data at $\sqrt{s}$ = 13 TeV. Figure~\ref{fig-1} (left) shows the inclusive $\psi$(2S)-over-J/$\psi$ cross section ratios as a function of \textit{p}$_{\mathrm{T}}$. In the measured range no significant \textit{p}$_{\mathrm{T}}$-dependence is found within the experimental uncertainties. Data are compared with NRQCD~\cite{MB_PRL106_2011} and ICEM~\cite{VC_PRD98_2018} model calculations and found to be in agreement within the theoretical and experimental uncertainties. The right panel of Fig.~\ref{fig-1} shows the first Run 3 results of $\sigma_{\psi(\mathrm{2S})}$ / $\sigma_{\textit{J}/\psi}$ versus \textit{p}$_{\mathrm{T}}$ at mid and forward \textit{y} in pp collisions at $\sqrt{s} =$ 13.6 TeV, compared to Run 2 results from TRD-triggered pp collisions at $\sqrt{s}=$ 13 TeV. It is worth noting that the large minimum-bias (MB) data sample already collected in Run 3 has allowed $\psi$(2S) measurement down to \textit{p}$_{\mathrm{T}}$ = 0. The $\psi$(2S)-over-J/$\psi$ cross section ratios obtained from Run 2 and Run 3 data are compatible within the uncertainties.
\begin{figure*}[ht]
\centering
\includegraphics[scale=0.25]{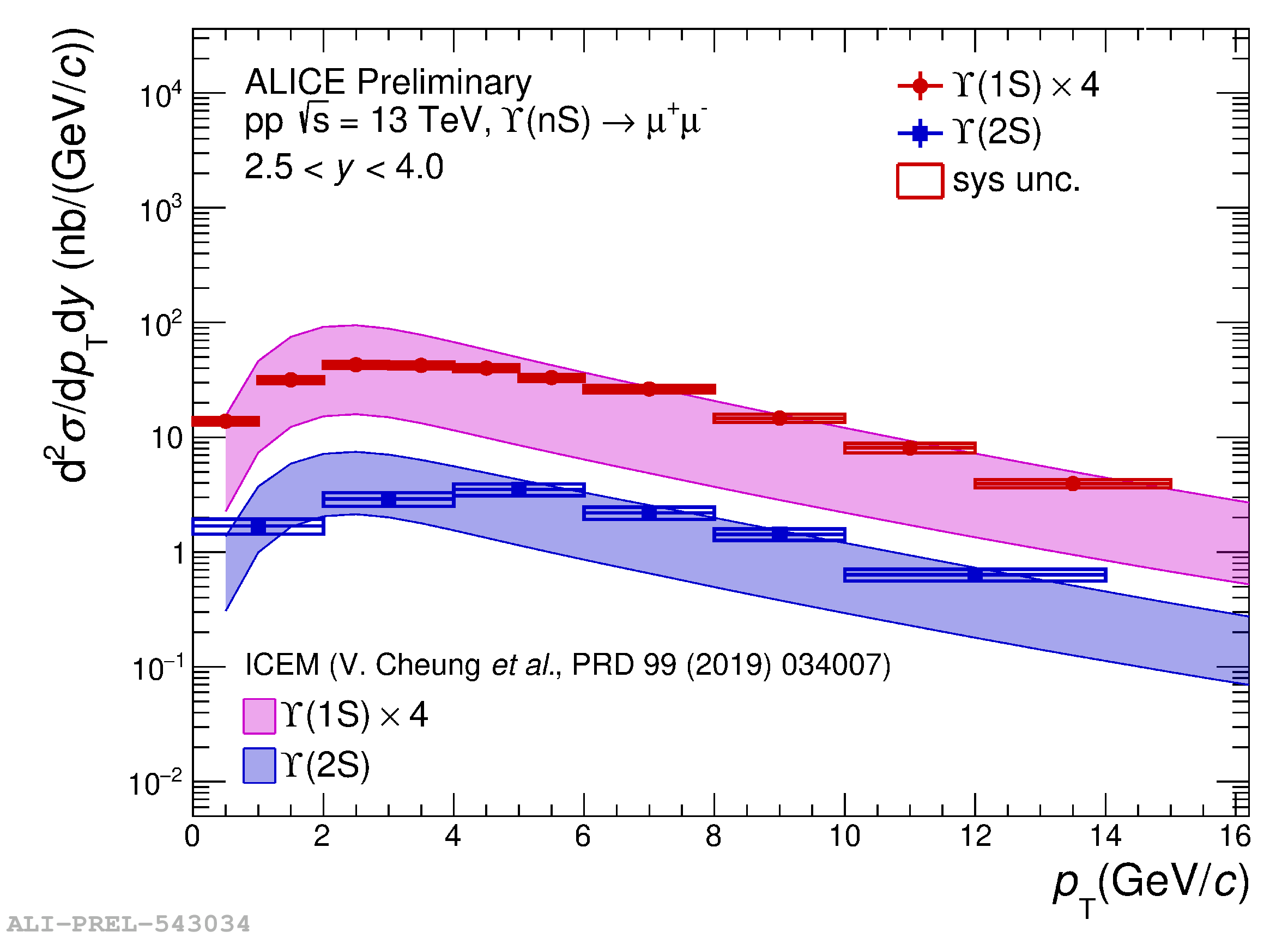}
\includegraphics[scale=0.24]{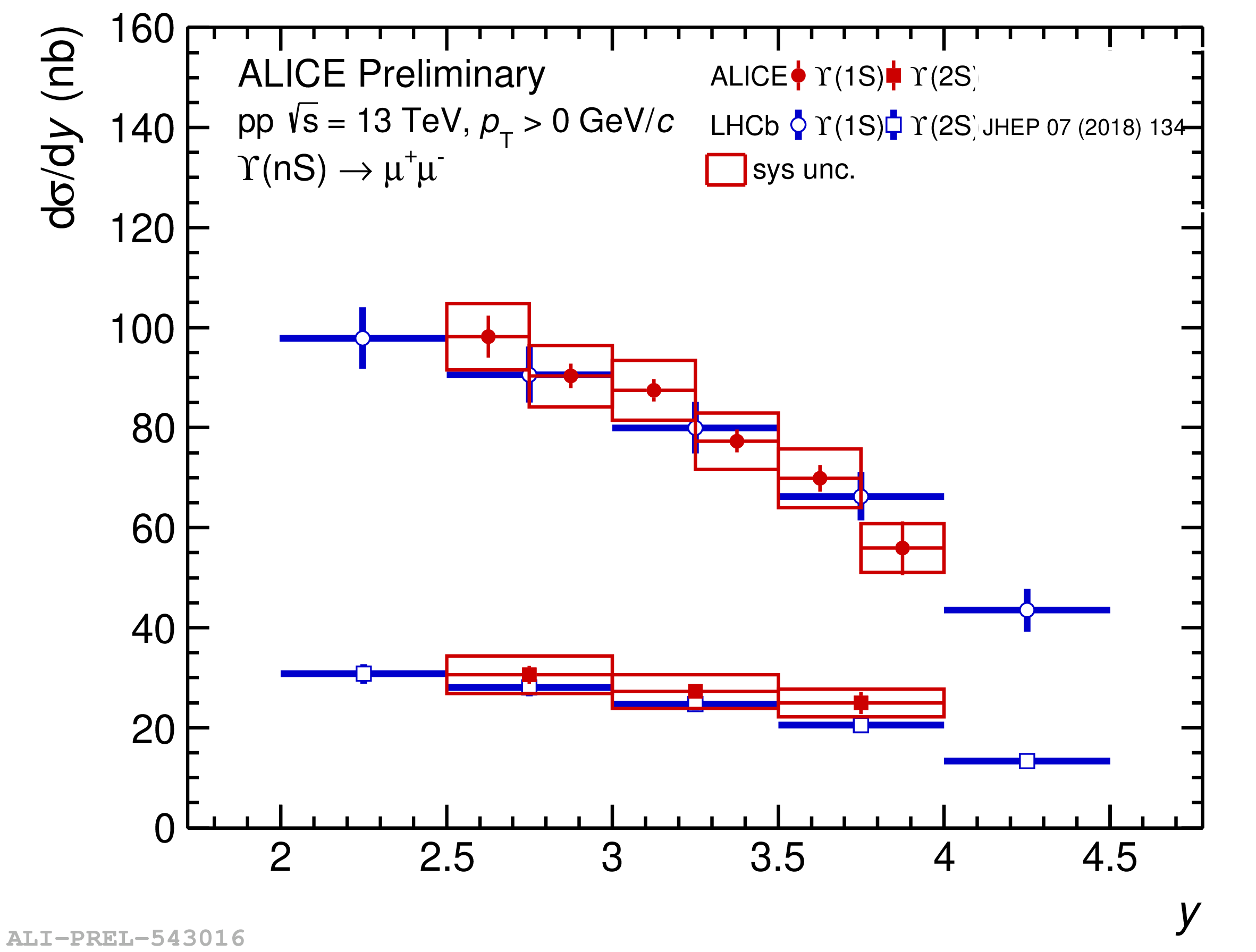}
\caption{(Left) Inclusive $\Upsilon$(1S) and $\Upsilon$(2S) cross sections, at forward \textit{y}, as a function of \textit{p}$_{\mathrm{T}}$, measured in pp collisions at $\sqrt{s}$ = 13 TeV. The data are compared with theoretical calculations from ICEM + FONLL~\cite{VC_PRD99_2019, FONLL}. (Right) Rapidity dependence of inclusive $\Upsilon$(1S) and $\Upsilon$(2S) cross sections, at forward \textit{y}, measured in pp collisions at $\sqrt{s}$ = 13 TeV. The results are compared with LHCb measurements at the same center-of-mass energy~\cite{LHCbRA_JHEP07_2018}.  }
\label{fig-2}       
\end{figure*}
In the bottomonium sector, \textit{p}$_{\mathrm{T}}$-differential inclusive $\Upsilon$(1S) and $\Upsilon$(2S) cross sections are measured, at forward \textit{y}, in pp collisions at $\sqrt{s}$ = 13 TeV and compared with ICEM+FONLL~\cite{VC_PRD99_2019, FONLL} model calculations, as shown in the left panel of Fig.~\ref{fig-2}. Within the theoretical and experimental uncertainties, data and model calculations are in agreement. In the right panel of Fig.~\ref{fig-2}, \textit{y}-differential cross sections for inclusive $\Upsilon$(1S) and $\Upsilon$(2S) are shown along with the comparison to LHCb~\cite{LHCbRA_JHEP07_2018} results at the same center-of-mass energy. ALICE and LHCb results are compatible within uncertainties.
\begin{figure*}[h]
\centering
\includegraphics[scale=0.24]{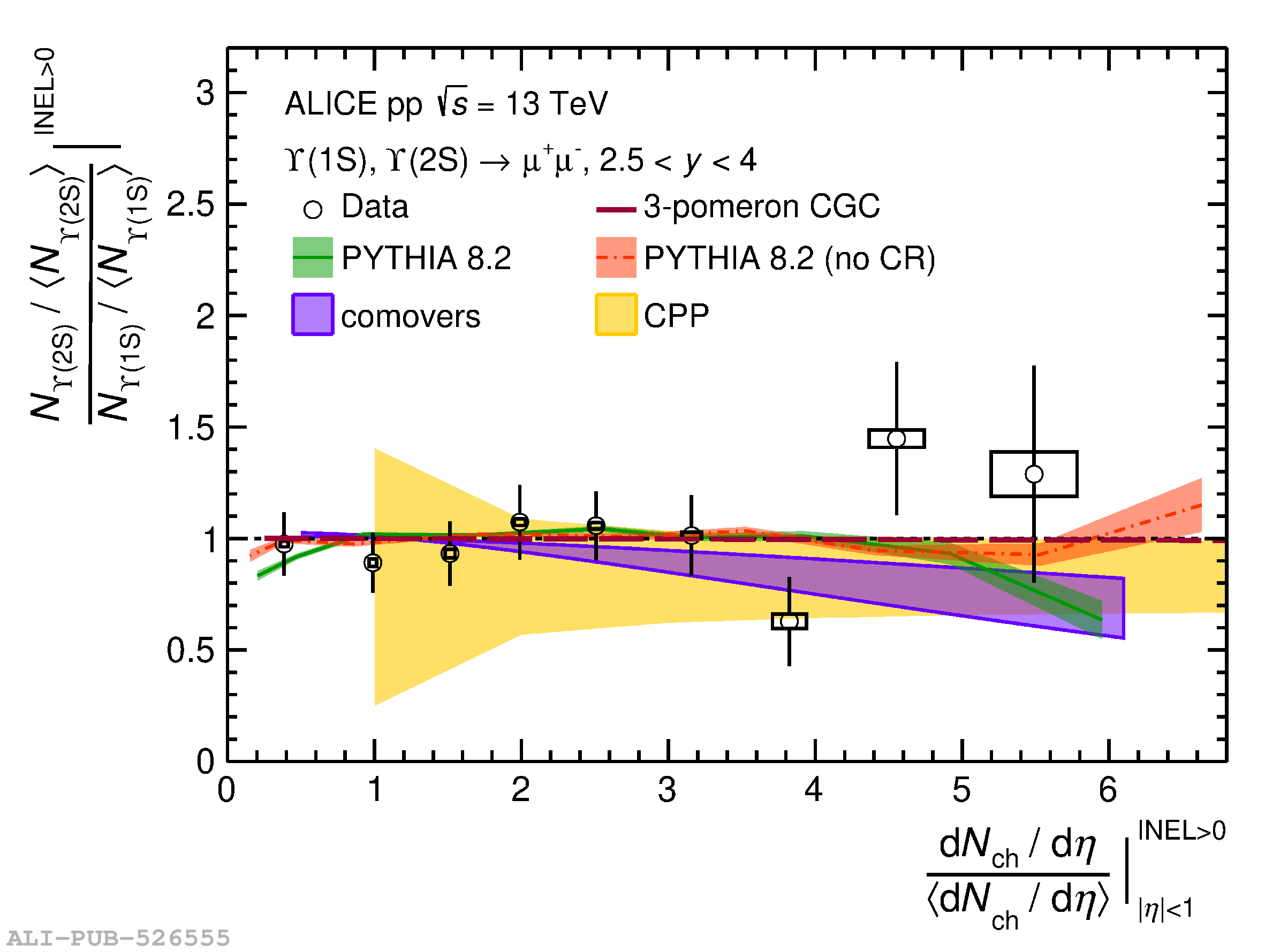}
\includegraphics[scale=0.24]{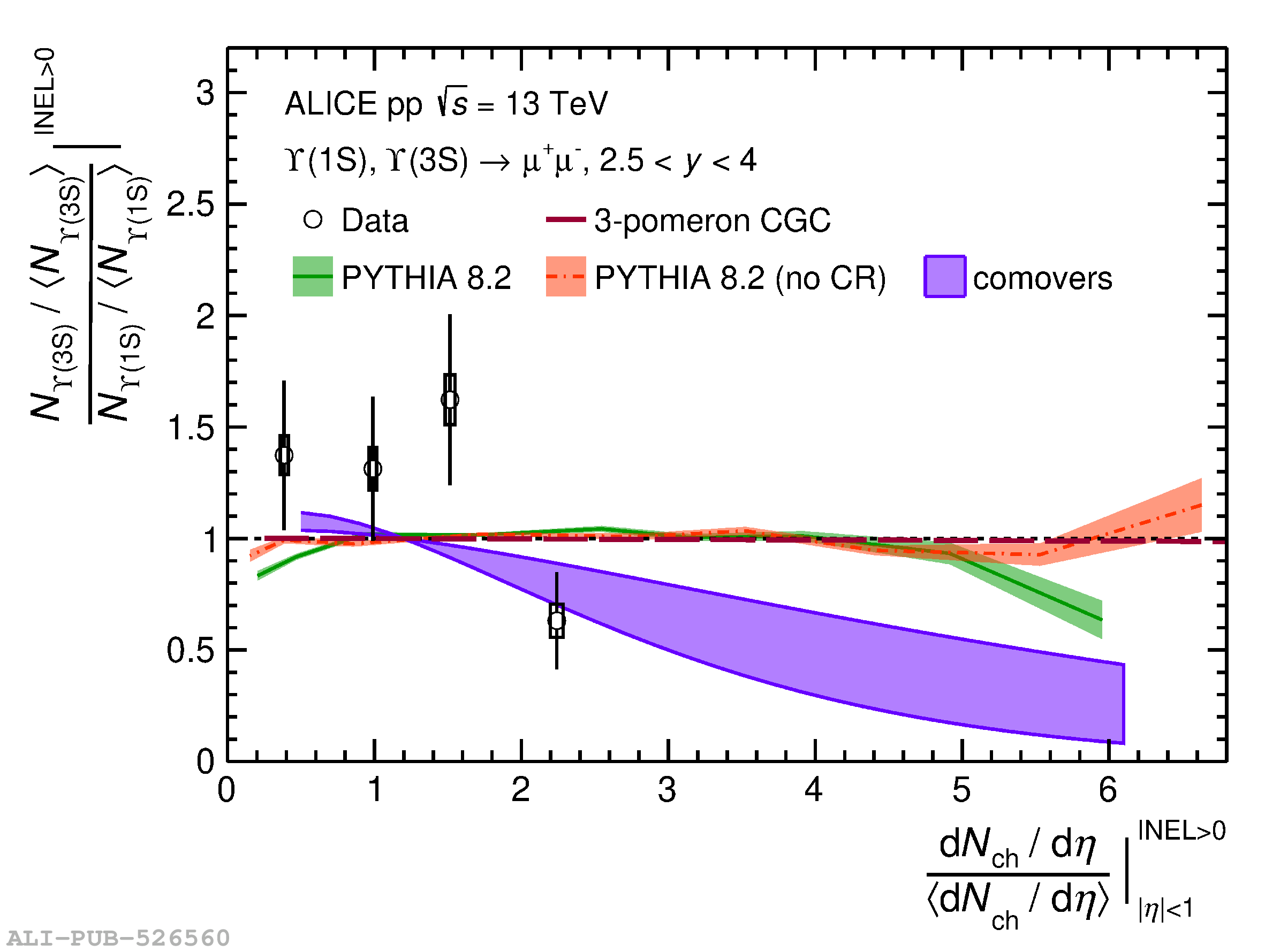}
\caption{Self normalized (left) $\Upsilon$(2S)-to-$\Upsilon$(1S) and (right) $\Upsilon$(3S)-to-$\Upsilon$(1S) yield ratios, at forward  \textit{y}, as a function of self-normalized charged-particle multiplicity at midrapidity in pp collisions at $\sqrt{s}$ = 13 TeV. The measurements are compared with PYTHIA 8.2 event generator, 3-pomeron-CGC, CPP, and comover models. Model descriptions can be found in~\cite{ALICE_arX2209}. }
\label{fig-3}       
\end{figure*}
As discussed previously, excited-to-ground state quarkonium yields versus multiplicity are sensitive to initial and final state interactions. Figure~\ref{fig-3} shows the self normalized $\Upsilon$(2S)-to-$\Upsilon$(1S) (left) and $\Upsilon$(3S)-to-$\Upsilon$(1S) (right) yield ratios as a function of the self-normalized charged-particle multiplicity at midrapidity, compared with model calculations that include, PYTHIA 8.2 w/ and w/o color reconnection (CR), 3-pomeron-CGC, coherent particle production (CPP), and comover approach. The essential features of these models are discussed in~\cite{ALICE_arX2209}. 
The measured ratios do not exhibit any strong multiplicity dependence within the current uncertainties, and they are compatible with all model calculations. 

\begin{figure*}[h]
\centering
\includegraphics[scale=0.22]{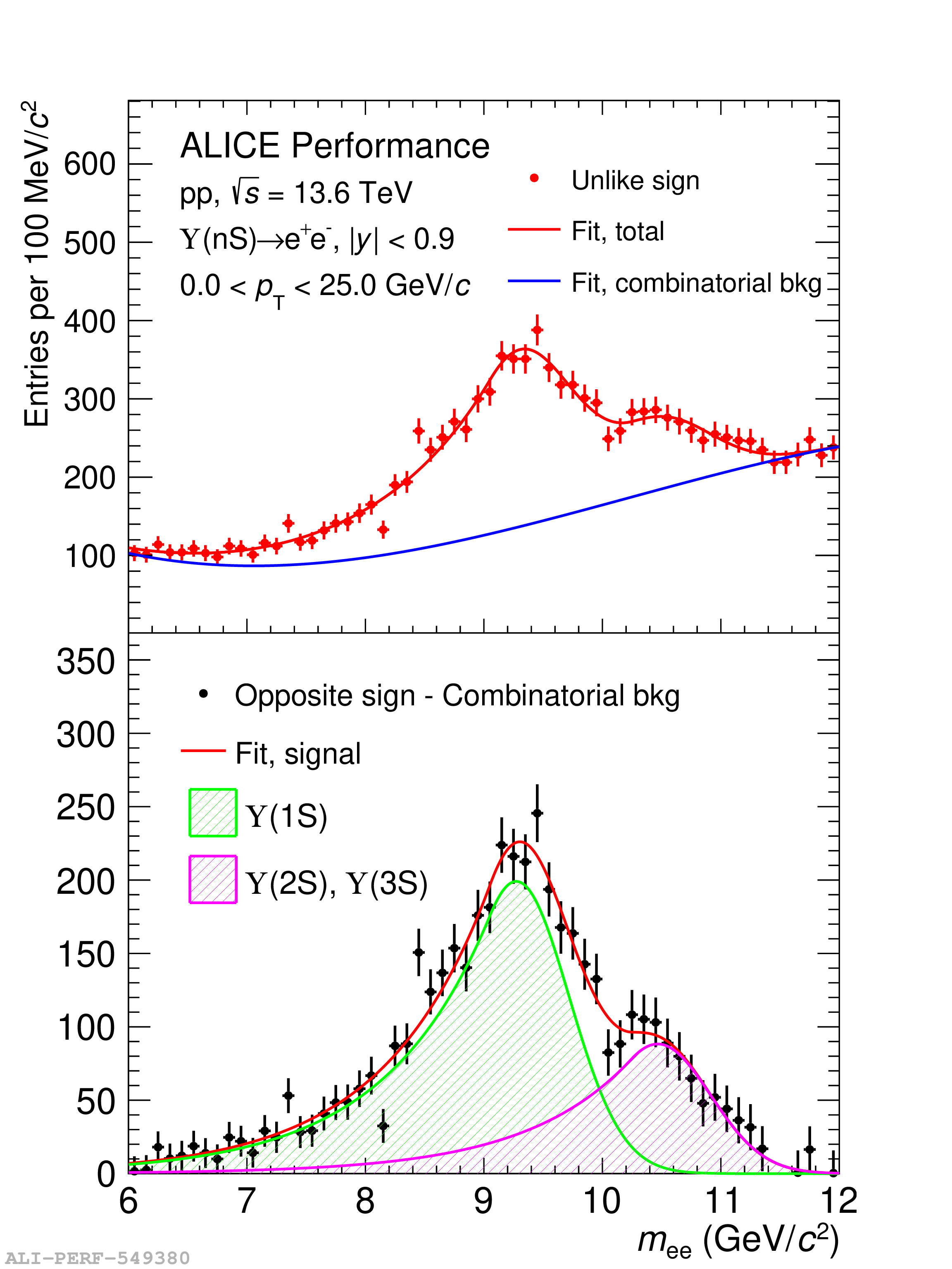}
\includegraphics[scale=0.22]{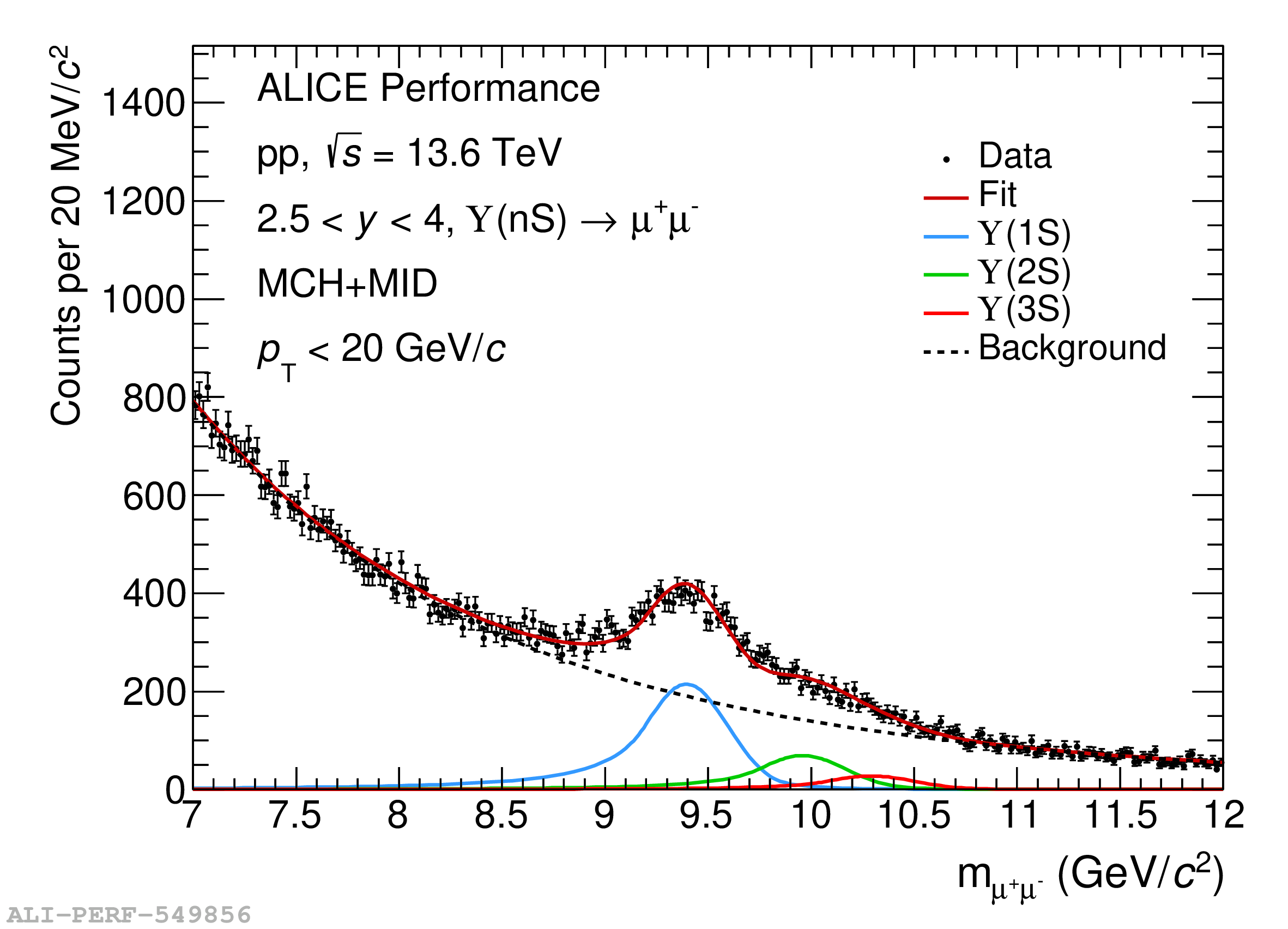}
\caption{Invariant mass distribution of $\Upsilon$(nS) obtained in the dielectron channel (left) at midrapidity and in the dimuon channel (right) at forward \textit{y} from Run 3 MB pp collisions at $\sqrt{s}$ = 13.6 TeV.}
\label{fig-4}       
\end{figure*}

Finally, few performance plots are shown that illustrate the ALICE potential for quarkonium measurements in Run 3, although, detector calibrations and alignment are still in progress. Figure~\ref{fig-4} (left) shows clear $\Upsilon$(nS) mass peaks in the dielectron decay channel, at midrapidity. At forward \textit{y}, the Fig.~\ref{fig-4} (right) shows that $\Upsilon$(nS) mass peaks are already visible. Thus, ALICE will be able to perform bottomonium measurements both at mid and forward \textit{y} in Run 3.

\vspace*{-0.50cm}
\section{Summary}
Quarkonium measurement is an important tool to learn about its production mechanism, and decipher the role of MPIs in small systems. 
These proceedings present first ever ALICE measurement of $\sigma_{\psi(\mathrm{2S})}$ / $\sigma_{\textit{J}/\psi}$ at midrapidity using TRD-triggered pp collisions at $\sqrt{s}$ = 13 TeV. Also shown are first Run 3 measurements for the same observable both at mid and forward \textit{y}. Inclusive $\Upsilon$(1S) and $\Upsilon$(2S) cross sections are presented as a function of \textit{p}$_{\mathrm{T}}$ and \textit{y} together with the multiplicity dependence of self normalized excited-to-ground $\Upsilon$(nS) state ratios. Given the current precision of the data, these ratios do not allow to conclude on the presence of final state effects. Lastly, first results on the ALICE performance for bottomonium detection in Run 3 are shown and discussed.

%
%
%

\end{document}